\documentclass[conference]{IEEEtran}
\IEEEoverridecommandlockouts
\usepackage{cite}
\usepackage{amsmath,amssymb,amsfonts}
\usepackage{algorithm}
\usepackage{algorithmic}
\usepackage{graphicx}
\usepackage{textcomp}
\usepackage{xcolor}
\usepackage{booktabs}

\def\BibTeX{{\rm B\kern-.05em{\sc i\kern-.025em b}\kern-.08em
    T\kern-.1667em\lower.7ex\hbox{E}\kern-.125emX}}
\begin{document}

\title{ An Actor-Critic-Based UAV-BSs Deployment Method for Dynamic Environments  \\
}

\author{\IEEEauthorblockN{Zhiwei Chen\IEEEauthorrefmark{2},
Yi Zhong\IEEEauthorrefmark{2},
Xiaohu Ge\IEEEauthorrefmark{2}\IEEEauthorrefmark{1}
}
\IEEEauthorblockA{\IEEEauthorrefmark{2}School of Electronic Information and Communications\\ Huazhong University of Science and Technology, Wuhan, China,\\
\{zhiwei\_chen, yzhong, xhge\}@hust.edu.cn}
\and
\IEEEauthorblockN{Yi Ma\IEEEauthorrefmark{3}
\IEEEauthorblockA{\IEEEauthorrefmark{3}Institute for Communication Systems (ICS)\\ University of Surrey, Guildford, England,\\
y.ma@surrey.ac.uk
}}
}
\maketitle
\begin{abstract}
  In this paper, the real-time deployment of unmanned aerial vehicles (UAVs) as flying base stations (BSs) for optimizing the throughput of mobile users is investigated for UAV networks. This problem is formulated as a time-varying mixed-integer non-convex programming (MINP) problem, which is challenging to find an optimal solution in a short time with conventional optimization techniques. Hence, we propose an actor-critic-based (AC-based) deep reinforcement learning (DRL) method to find near-optimal UAV positions at every moment. In the proposed method, the process searching for the solution iteratively at a particular moment is modeled as a Markov decision process (MDP). To handle  infinite state and action spaces and improve the robustness of  the decision process, two powerful neural networks (NNs) are configured to evaluate the UAV position adjustments and make decisions, respectively. Compared with the heuristic algorithm, sequential least-squares programming and fixed UAVs methods, simulation results have shown that the proposed method outperforms  these three benchmarks in terms of the throughput at every moment in UAV networks.
\end{abstract}
\begin{IEEEkeywords}
UAV deployment, deep reinforcement learning, throughput maximization, dynamic user, actor-critic.
\end{IEEEkeywords}

\section{Introduction}

Unmanned aerial vehicles  have been proposed in future wireless networks to act as mobile base stations (UAV-BSs) in order to offer emergency communication or provide wireless services to underserved areas \cite{Zeng:2016tz}, \cite{Mozaffari:2019hh}. Compared with  ground base stations (GBSs), UAV-BSs can be deployed flexibly, which will increase the probability of establishing  line-of-sight (LOS) links to moving users. Hence, the deployment of UAV-BSs is one of the key design considerations in future heterogeneous wireless networks for capacity maximization, smart city, mobile edge computing and autonomous vehicular networks \cite{7886285}.

As a result, the  deployment of UAV-BSs has received significant attentions. For instance, the optimal deployment and mobility of multiple UAVs for  data collection from IoT sensors was researched \cite{8038869}. Furthermore, the optimal altitude, which enables one UAV to achieve maximum coverage, was investigated by \cite{AlHourani:2014gj}. The authors maximized the minimum throughput over all ground users by optimizing the UAV trajectory and power control joint with user communication scheduling and association \cite{Wu:2018ev}, where the non-convex optimization problem was solved via applying block coordinate descent and successive convex optimization techniques. The article \cite{Liu:2018ej}  utilized the deep deterministic policy gradient algorithm (DDPG) \cite{lillicrap2015continuous} to maximize the UAV power with  consideration for fairness, communication coverage and connectivity. The evolutionary algorithm was adopted to find the optimal deployment of UAV-BSs for disaster relief scenarios in \cite{6881196}.

However, only stationary scenarios are configured for users in the above studies for UAV-BS deployment optimizations. In real-life scenarios users often randomly move, which results in users being at different positions in different time slots and difficulties in the evaluation of the random network performance\cite{7399689,7015548}. To improve the performance of  UAV networks, UAV-BSs have to quickly adjust their positions considering different locations of users. In this case, a rapid UAV-BS deployment method considering moving users needs to be investigated for dynamic environments.

UAV-BS deployment in dynamic environments is usually formulated as a time-varying mixed-integer
non-convex programming (MINP) problem, which is a type of NP-hard problem. For this NP-hard problem, heuristic algorithms  are applied to find the near-optimal solution with computation power. Joint optimization and evaluation of the computation and communication power in cellular networks was investigated in \cite{7054598},\cite{6416884}.  Meanwhile the  heuristic algorithms is often fall into the optimal local solution. Hence, an actor-critic (AC) based UAV-BS deployment method based on the deep reinforcement learning (DRL) framework is proposed to improve the real-time network throughput. The main contributions and innovation of this paper are summarized as follows:
{\begin{enumerate}
  \item To avoid solving the MINP problem directly, the process of finding the optimal position of the UAV-BS is formulated as a Markov decision process (MDP). To search for the best policy function of the MDP where the state and action space are continuous position values generated by the UAV-BSs and users, an AC-based DRL method is adopted in dynamic environments. Compared with the deep Q learning method, the AC-based method can avoid to fall into the optimal local solutions which is caused by the Bellman optimal equation of state-action value function.
  \item Instead of simply setting the reward function empirically, a mathematical expression of the immediate reward function is derived considering the MINP optimization objective. The mathematical expression of reward function ensures the convergence of the actor and critic neural networks (NNs).
  \item Based on the proposed AC-based method, the UAV-BS agent finally learn the cooperation among UAV-BSs and effectively reduce mutual interference. Simulation results indicate the proposed method has a good generalization when users are located in different probability distributions.
\end{enumerate}}
The rest of this paper is organized as follows. In Section II, the system model and problem formulation are introduced. The AC-based method is depicted for UAV-BS deployment in Section III. Simulation results are presented and analyzed in Section IV. Finally, the conclusions are given in Section V.

\begin{figure}[htbp]
\centerline{\includegraphics[width=0.40\textwidth, height=0.3\textwidth]{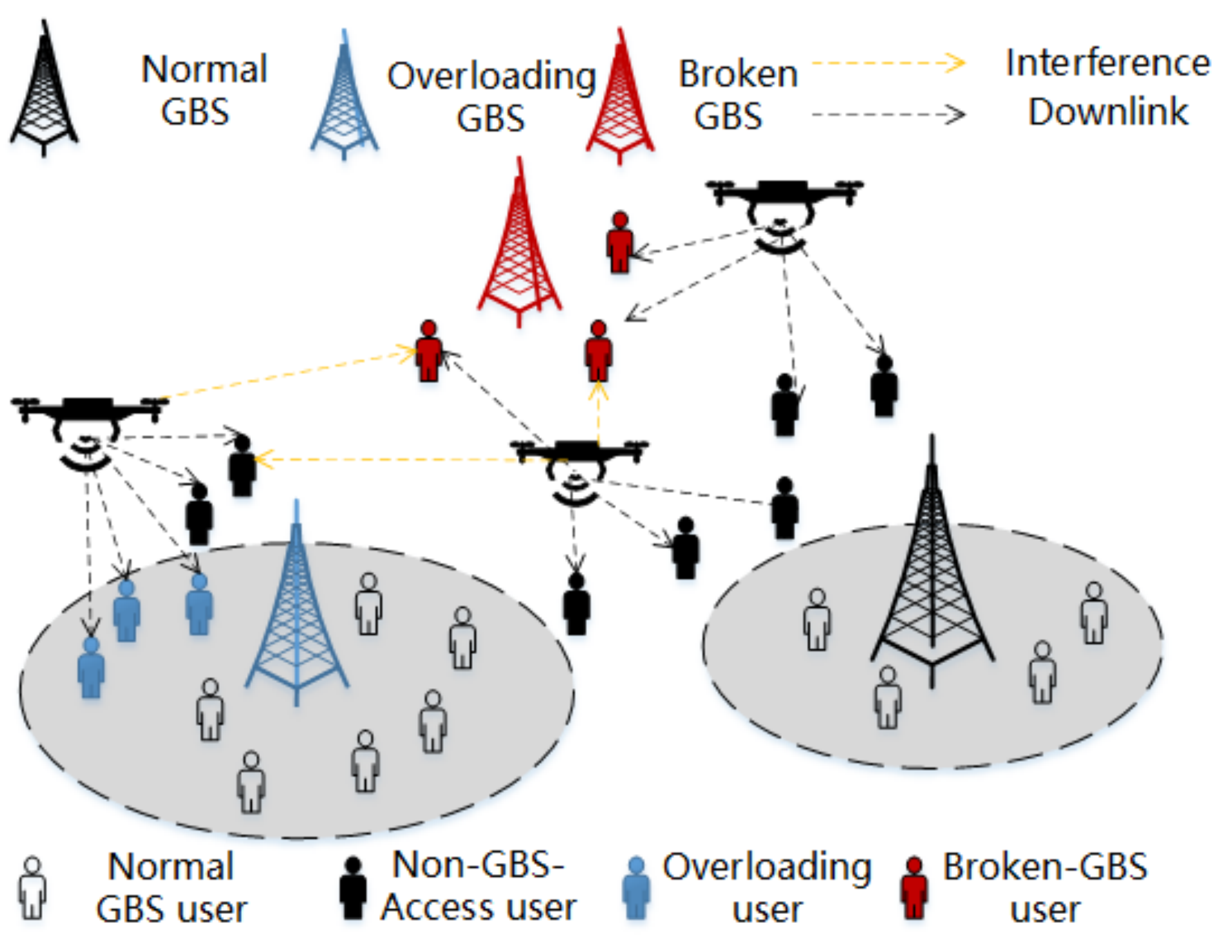}}
\caption{UAV networks.}
\label{fig}
\end{figure}

\section{System Model}
Two important reasons for adopting UAV to assist current communication systems are listed as follows: one reason is that UAV-BSs can be used for emergency communications; the other reason is that UAV-BSs fly to area without ground BSs coverage to provide wireless services. A UAV network is configured in Fig. 1, where a set \(\mathcal{G}\) includes a number of \(G\) ground BSs, a set \(\mathcal{K}\) includes a number of \(K\) users, a set \(\mathcal{P}\) includes a number of \(P\) UAV-BSs. The downlink (DL) is denoted as the wireless link from a UAV-BS \(p\) to a user \(k\). We consider the sub-6 GHz band for the UAV-BS data links and all the UAV-BSs share the same frequency band for wireless communications. The total bandwidth is equally divided among the associated users.  All users are randomly located and equipped with the single antenna. The time division multiplexing scheme is adopted in UAV networks. Each user is allowed to access only one UAV-BS in one time slot. For ease of exposition, we assume that the UAV-BSs use a different frequency  from the ground BSs to avoid interference between  UAV-BSs and ground BSs.

Without loss of generality,  the 3D cartesian coordinate of each UAV-BS \(p\in \mathcal{P} \) and user \(k\in \mathcal{K} \) are \((x_p(t),y_p(t),H)\) and \((x_k(t),y_k(t),0) \), where $x_p(t)$ is the UAV-BS $x$ axis value and $y_p(t)$ is the UAV-BS $y$ axis value at the time slot $t$ with the altitude value $H$, $x_k(t)$ is the user $x$ axis value and $y_k(t)$ is the user $y$ axis value at time slot $t$. All UAV-BSs are assumed to fly at a fixed altitude \(H\) above the ground and  the horizontal coordinates of each user and  the horizontal coordinates of UAV-BS are changed with time. The reason why the UAV fly at a fixed altitude is that  searching the best UAV-BSs 3-D positions could be divided into two stages. This work focuses on the first stage, which is letting the UAV-BSs  in the right horizontal position at a short time. The Euclidean distance between  a UAV-BS \(p\) and a user \(k\) at a time slot \(t\) can be expressed as \(r_{p,k}(t)=\sqrt{(x_k(t)-x_p(t))^2+(y_k(t)-y_p(t))^2+H^2},k\in \mathcal{K} , p \in \mathcal{P}\).

\subsection{Channel Model}
The channel gain \(g_{k,p}\) between
a UAV-BS \(p\) and a user \(k\) is composed of a line-of-sight component \(g_{k,p}^{LOS}\) and a non-line-of-sight (NLOS) component \(g_{k,p}^{NLoS}\).
 the \(g_{k,p}^{LOS}\) and \(g_{k,p}^{NLOS}\) can be  given as follows\cite{7037248}
$$
\left\{\begin{aligned} g_{k, p}^{\mathrm{LOS}} &=\left(\frac{4 \pi f}{v}\right)^{2} \mu_{\mathrm{LOS}} r_{k, p}^{-\alpha_{\mathrm{LOS}}},  \\ g_{k, p}^{\mathrm{NLOS}} &=\left(\frac{4 \pi f}{v}\right)^{2} \mu_{\mathrm{NLOS}} r_{k, p}^{-\alpha_{\mathrm{NLOS}}} , \end{aligned}\right.
\eqno(1)
$$
Where \(f\) is the carrier frequency, \(v\) is the light speed, \(\alpha_{LOS}\) and \(\alpha_{NLOS}\) is the path loss exponent in the LOS and NLOS transmission conditions, respectively. Compared with the influence induced by the NLOS transmission, the impact of multi-path fading can be neglected \cite{7037248}. \(\mu_{\phi}(\phi \in\{\mathrm{LOS}, \mathrm{NLOS}\})\) is the attenuation factor. The user received power \(P_{rec}(r)\) can be written as
$$
P_{rec}(r)=\left\{\begin{array}{l}{P_{V} g_{k, p}^{\mathrm{LOS}} , LOS\ link}, \\ {P_{V} g_{k, p}^{\mathrm{NLOS}}, NLOS\ link}, \end{array}\right. \eqno(2)
$$
where \(P_{V}\) is the UAV-BS transmit power. Assuming that all  UAV-BSs have the same  \(P_{V}\). Here, the probability of LOS connection depends on the different environments, density, height of  buildings and the elevation angle between  users and  UAV-BSs. The LOS probability can be expressed as follow \cite{AlHourani:2014gj}

$$
P_{LOS}=\frac{1}{1+C\exp (-B[\theta-C])}, \eqno(3)
$$
where $B$ and $C$ are constants which depend on the environments. \(\theta\) is the elevation angle. For the user \(k\) and UAV-BS \(p\), \(\theta_{p,k}\) is
$$
\theta_{p, k}=\frac{180}{\pi} \times \sin ^{-1}\left(\frac{H}{r_{p,k}}\right). \eqno(4)
$$
The probability of NLOS link is \(P_{NL O S}=1-P_{L O S}\). Furthermore, the average received power of the user \(k\) served by the UAV-BS \(p\) can be expressed as
$$
P_{k, p}=P_{L O S} P_{rec}(r) +P_{N L O S} P_{rec}(r).  \eqno(5)
$$

\subsection{Problem Formulation}
Considering that all UAV-BSs share the same frequency band, the interference among UAV-BSs can't be ignored in UAV networks. The signal-to-interference-plus-noise ratio (SINR) at a user equipment (UE) \(k\) from UAV-BS \(p\) at time $t$ is
$$
SINR_{k,p,t}=\frac{P_{k,p,t}}{\sum_{j \in \mathcal{P} \backslash p } P_{k,j,t}+\sigma^{2}} ,
 \eqno(6)
$$
where \(\sigma^2\) is the noise power at the user side which is considered to be the additive white Gaussian noise (AWGN). \(P_{k,j,t}\) is the received power of the user \(k\) served by the UAV-BS \(j\) at time \(t\). The transmission rate of a UE \(k\) at time \(t\) can be expressed as:
$$
R_{k,p,t}=a_{k,p,t}  \log _{2}\left(1+{SINR_{k,p,t}}\right), k \in \mathcal{K}, p \in \mathcal{P} , \eqno(7)
$$
where \(a_{k,p,t}\) is the indicator factor. \(a_{k,p,t}=1\) means a user \(k\) is served by a UAV-BS \(p\) at time \(t\), otherwise, \(a_{k,p,t}=0\).

In this paper, a period of time \(T\) is discretized into
multiple time slots and  users in each time slot are assumed
to be in  same positions. The UAV-BSs deployment aims to maximize the real-time throughput by moving to the proper positions at every time slot.  Mathematically, this  problem can be formulated as
$$
\max   \sum_{k \in \mathcal{K}} \sum_{p \in \mathcal{P}} {a_{k,t}R_{k,p,t}}, \forall t\in T ,   \eqno(8)
$$
$$
\begin{aligned}
s.t.\  a_{k,p,t}&=\mathbf{1}_{[0,\infty ]}[R_{k,p,t}-\Gamma]\times \mathbf{1}_{[0,\infty ]}[\delta-r_{k,p,t}],\\
 & \forall (k,p) \in \mathcal{K} \times \mathcal{P},\forall t\in T ,
\end{aligned}
\eqno(9)
$$
$$
\mathbf{1}_{[0,\infty]}[n]=\left\{\begin{array}{ll}{0,} & {
n<0} \\ {1,} & {n \geq 0}\end{array}\right .\eqno(10)
$$
$$
x_{k,t},x_{p,t}\in[0,A_X],\forall t\in T , \eqno(11)
$$
$$
y_{k,t},y_{p,t}\in[0,A_Y],\forall t\in T , \eqno(12)
$$
where \(a_{k,p,t}=1\) when the  distance  between  UAV-BS \(p\) and user \(k\) is shorter than the UAV-BS \(p\) communication  distance \( \delta\) and the rate of user \(k\) is larger than the communication rate threshold \(\Gamma\), otherwise, \(a_{k,p,t}=0\). (11) and (12) imply that UAV-BSs and  users can only move in a given \(A_x\times A_y\) rectangular area.

Since the problem at time slot \(t\) is a mixed-integer
non-convex programming (MINP) problem which is challenging to solve due to these two reasons as follow: first, \(R_{k,p,t}\) is a non-convex  function respect to the distance between  UAV-BSs and users \(r_{k,p,t}\); Second, the  variable \(a_{k,p,t}\) is binary. Conventional optimization technique like the convex optimization can not used directly for this problem due to the high complexity. In general, heuristic algorithms are adopted to find a near-optimal solution. However, the heuristic algorithm has to be re-run as long as the user locations have been changed, which causes a high computational overhead. Different from heuristic algorithms, the DRL agent can learn the  policy of UAV-BS position adjustments based on the continuous interactions with the environment and then save the  policy as the deep neural network weights. Considering the continuous position values generated by the UAV-BSs and users, AC-based  method is adopted to solve this UAV-BS deployment optimization problem.
\section{AC-based Deployment Optimization of UAV-BSs}
In this section, we propose an AC-based method combining
the advantages of deep Q learning  and policy gradient
 \cite{mnih2015human}, where each UAV-BS can find  intelligently
and quickly the target regions in every time slot.

\subsection{Preliminaries}
Based on a standard reinforcement learning setting, an agent interacts with a system environment in discrete epochs. The agent observes a state \(s_i\), executes a action \(a_i\) and obtains a reward \(r_i\) at a epoch \(i\). As usual,  the episode is written as \((s_0,a_0,s_1,a_1,...,s_i,a_i,s_{i+1})\) after several multiple actions and state transitions. The reinforcement learning objective is to find a good policy function \(\pi(a_i)\) that maps a state to an action for maximizing the discounted cumulative reward of a MDP, which is also called the value function. The discounted cumulative reward can be denoted as
$$
V^{\pi}(s)=\mathbb{E}\left[\sum_{i=0}^{\infty} \gamma^{i} r(s_i, a_i) | s_0={s}\right],
\eqno(13)
$$
where \(r(s_i,a_i)\) is the immediate reward, \(\gamma^{i} \in [0, 1]\) is the discount factor at the epoch i and \(\mathbb{E}[\cdot]\) is the expectation operation. The state \(s_i\) and action \(a_i\) is usually combined as a state-action pair \((s_i,a_i)\). Hence, the value function can be replaced as a Q-value function \(Q(s_i,a_i)\). The Q-value function can be expressed as
$$
\begin{aligned}
Q(s_i, a_i)&=r(s_i, a_i) \\
&+\gamma \sum_{s_{i+1} \in \mathcal{S}} \sum_{a_{i+1} \in \mathcal{A}} P_{s_{i} s_{i+1}}(a_i) Q( s_{i+1}, a_{i+1}),\\
\end{aligned}
\eqno(14)
$$
where \(P_{s_{i} s_{i+1}}\) is the transition probability from the state \(s_i\) to the
next state \(s_{i+1}\). \(\mathcal{S}\) and \(\mathcal{A}\) are the state space and action space, respectively. Considering that the state transition probability is unknown in  dynamic environments, the deep Q learning algorithm is proposed by adopting  a deep neural network (DNN) to approximate the Q-value function. The deep Q network (DQN) is trained by minimizing the following loss function
$$
L(\theta^Q)=\ E[(y_i-Q(s_i,a_i|\theta^Q))^2], \eqno(15)
$$
where \(\theta^Q\) is the weight vector of the DQN. \(y_i\) is the target value, which can be estimated by a temporal difference approach \cite{sutton1998introduction}. Hence, \(y_i\) is denoted by
$$
y_i= r(s_i,a_i)+\gamma^i Q(s_{i+1}, a_{i+1})|\theta^Q). \eqno(16)
$$

However, the deep Q learning algorithm only works for problems with a discrete action space. To apply the DRL for the high dimensional continuous problem, the AC method has been proposed. In the AC method, the action is generated by the policy function  which can be approximated by a DNN.  The policy DNN  are updated by gradient ascent and the loss of the policy DNN is the expectation of discounted cumulative reward of multiple episodes, which is written as follows:
$$
\overline{R}_{\theta^\pi}=E_{\tau \sim p_{\theta^\pi}(\tau)}[R(\tau)]
\approx \frac{1}{N} \sum_{n=1}^{N} R(\tau^n) , \eqno(17)
$$
where the \(p_\theta (\tau)\) is the probability of the episode \(\tau\) showing up at the policy DNN with the weight \(\theta^\pi\). In the AC method, \(R(\tau)\) can be estimated by the Q-value  produced by the DQN. Therefore, the  actor DNN gradient updating expression is
$$
\theta^\pi \leftarrow \theta^\pi+\eta \nabla \overline{R}_{\theta^\pi}, \eqno(18)
$$
where the  \(\nabla \overline{R}_{\theta^\pi}\) can be calculated by:

$$
\begin{aligned}
\nabla \bar{R}_{\theta^\pi} &\approx \mathbb{E}[\nabla_{\boldsymbol{\theta}^{\pi}} Q(s,a|\theta^Q)|_{s=s_i,a=\pi(s_i|\theta^\pi)}]\\
&= \mathbb{E}[\nabla_a Q(s,a|\theta^Q)|_{s=s_i,a=\pi( s_i )} \\
&\cdot \nabla_{\theta^{\pi}}\pi(s|\theta^\pi)|s=s_i].
\end{aligned}
\eqno(19)
$$

  Different from the  policy gradient method, the AC method adopts the output value of DQN  as a loss function to improve the robustness of policy function. Compared with both the deep Q learning and  policy gradient methods, the AC method is difficult to train and converge.

\subsection{AC-based UAV-BS Deployment Method  }
Considering the time-varying MINP problem in maximizing
the real-time throughput of UAV networks,
we transfer the MINP problem into an MDP problem at every time slot, which can
be solved by the DRL algorithm rather than solving the MINP problem directly and
violently. At every time slot $t$, the DRL agent can output the change of position
of UAV-BS in multiple epochs iteratively. To begin with, the state space, action space and the reward function are denoted
in the following:

1) \textbf{State space}: \(s_i(t)\) is the state at an epoch \(i\) when the time slot is \(t\). The state  information includes the  position of each user,  position of each UAV-BS and each  user association with UAV-BSs. There are \(K\times 3+ P \times 2 \) elements in the state. The first group is \([u_{i,k}(t),\forall k\in \mathcal K ]\) representing the user horizontal coordinate positions. The second group is \([ q_{i,p}(t),\forall p\in \mathcal P ]\) which denotes the UAV-BS positions. The third part is \([c_{i,k}(t),\forall k\in \mathcal U ]\) showing the user association with UAV-BSs, where the value is the number of UAV-BSs.
Hence, the \(s_i(t)\) is expressed as
$$
\begin{aligned}
  s_i(t)=[& u_{i,0}(t), ... , u_{i,K-1}(t);\\
               & q_{i,0}(t), ... , q_{i,P-1}(t);\\
               & c_{i,0}(t), ... , c_{i,K-1}(t);].
\end{aligned}
\eqno(20)
$$

2) \textbf{Action space}: \(a_i(t)\) is the action at a epoch \(i\) when the time slot is \(t\). \(a_i(t)=[u_{i+1,p}(t)-u_{i,p}(t)]\) is the change of horizontal coordinates of the UAV-BS positions between the epoch \(i\) and epoch \(i+1\). Hence, \(a_i(t)\) can be expressed as
$$
a_{i}(t)=[\Delta x_{i,0}(t),...,\Delta x_{i,P}(t);\Delta y_{i,0}(t),...\Delta y_{i,P}(t)].
\eqno(21)
$$

3) \textbf{reward}: The MINP objective should be the  cumulative  reward of  a episode at every time slot. \(\sum_{k \in \mathcal{K}} \sum_{p \in \mathcal{P}} {a_{k,t}R_{k,p,t,I}}\) is the throughput of UAV network at the last epoch \(I\) when the time slot is \(t\), i.e.
$$
\sum_{k \in \mathcal{K}} \sum_{p \in \mathcal{P}} {a_{k,t}R_{k,p,t,I}} =\sum_{i=1}^{I} r_i(s_i,a_i).
\eqno(22)
$$
 The throughput of UAV network at the initial epoch is considered to be zero at every time slot. Hence, (22) can be rewritten as
$$
\begin{aligned}
    \sum_{i=1}^{I} r_i(s_i,a_i)&=\sum_{k \in \mathcal{K}} \sum_{p \in \mathcal{P}} {a_{k,t}R_{k,p,t,I}}-\sum_{k \in \mathcal{K}} \sum_{p \in \mathcal{P}} {a_{k,t}R_{k,p,t,0}}  \\
   &=\sum_{i=1}^{I}(\sum_{k \in \mathcal{K}} \sum_{p \in \mathcal{P}} {a_{k,t}R_{k,p,t,i}}  \\
   &-\sum_{k \in \mathcal{K}} \sum_{p \in \mathcal{P}} {a_{k,t}R_{k,p,t,i-1}})  \\
\end{aligned}
,
\eqno(23)
$$
where \(\sum_{p \in \mathcal{P}} {a_{k,t}R_{k,p,t,0}}=0\). From (23), the immediate reward is derived by
$$
r_i(s_i,a_i)= \sum_{k \in \mathcal{K}} \sum_{p \in \mathcal{P}} {a_{k,t,i}R_{k,p,t,i}}-\sum_{k \in \mathcal{K}} \sum_{p \in \mathcal{P}} {a_{k,t}R_{k,p,t,i-1}}.
\eqno(24)
$$
Based on (24), the change of throughput between the epoch \(i\) and \(i-1\) is considered as the immediate reward at epoch $i$.

 A centralized ground base station is responsible for training process and the inference process is executed by each UAV-BS independently. The UAV-BSs are also connected to ground base stations to get core network service. The centralized trainer collects experiences  in dynamic environments and trains the actor and critic DNNs. Each UAV-BS as an agent equipped with the same actor DNN adjusts their position independently. The Markov chain \((s_0(t),a_0(t),...,s_i(t),a_i(t),s_{i+1}(t))\) denotes a process searching for the optimal position iteratively at a time slot \(t\). Given the initial positions of the UAV-BSs and users, the UAV-BS as an agent  outputs a  position change \(a_i(t)\) step by step and arrives at the target position in the end. The MINP problem constraints are included in the environment configuration.

The main steps of training process and decision process are presented in Algorithm 1. Without loss of generality, all users and UAV-BSs are randomly located and the weights of both actor and critic DNN are random. Each UAV-BS with state \(s_i(t)\) executes the action \(a_i(t)\) to move to a new position and obtains an immediate reward \(r_i(s_i(t),a_i(t))\) as well as a new state \(s_{i+1}(t)\) in the epoch \(i+1\). After executing above steps, the new experience \(( s_i(t); a_i(t); r_i(t); s_{i+1}(t))\) has been collected into the replay buffer \(B\) (i.e. the step 8-10). A mini-batch of experiences with size $N$ is randomly sampled from \(B\). The  DQN target value is calculated  by (15), i.e. the step 11-12. The step 14 is used to  update the weights \(\theta^Q\) of the critic DNN considering the loss function (14) and the weights of actor DNN are updated via (17), where the gradient is calculated by (18). Besides, the target actor DNN and target critic DNN are updated in every \(L\) epoch, i.e., the step 13-14. In the decision process, the UAV-BS agent interacts with the dynamic environment  as long as the user locations are changed. At every time slot \(t\), a Markov chain is formed. In the last epochs of the entire Markov chain, the action value converges to 0 which means UAV-BS no longer change its position. And then, the state with the maximum throughput is selected for each episode.

\begin{algorithm}[h]
\caption{AC-based UAV deployment}
\begin{algorithmic}[1]
\STATE \textbf{Training Process}
\STATE Randomly initializes the critic DNN \(Q(s,a|\theta^{Q})\) and actor DNN \(\pi(s|\theta^{\pi})\) with weights \(\theta^Q\) and \(\theta^{\pi}\);
\STATE Initializes target critic DNN \(Q^{'}(s,a|\theta^{{Q}^{'}})\) and the target actor DNN \(\pi(s|\theta^{\pi^{'}})\) with weights \(\theta^{Q^{'}}=\theta^{Q} \) ,\(\theta^{\pi^{'}}=\theta^{\pi}\);
\STATE Initializes the replay buffer $B$;
\FOR {time slot :=1, ... , $N$}
\STATE Initializes the environment and receives an initial state \(s_1\);
\FOR {epoch:=1, ... , $I$}
\STATE \(\boldsymbol{a}_{i}=\pi\left(\boldsymbol{s}_{i}\right)+\mathcal{N}\);
\STATE Executes \(a_i\) and obtains the new state \(s_{i+1}\), reward \(r_{i}\);
\STATE Stores the transition sample \((s_i,a_i,r_i,s_{i+1})\) into $B$;
\STATE Samples a mini-batch of \(H\) samples \((s_j,a_j,r_j,s_{j+1})\) from $B$;
\STATE Calculates the target value \(y_{j}\) by (15);

\STATE Updates the  weights of critic DNN  \(\theta^{Q}\) by minimizing the loss by (14):

\STATE Updates the  weights \(\theta^{\pi}\) of actor DNN by \(\theta \leftarrow \theta+\eta \nabla \overline{R}_{\theta}\) , where the gradient is calculated by (18):

\STATE In every L epoch, update the corresponding target DNNs:
\ENDFOR
\ENDFOR
\STATE \textbf{Decision Process}

\WHILE {the initial state \(s_0(t)\) updates }
\FOR {epoch := 1, ... ,I}
\STATE UAV as a DRL agent executes a action \(a_i\), obtains new state \(s_{i +1}\), reward \(r_i\).
\ENDFOR
\STATE Find the \(s_i\) with maximum throughput in one episode \([s_1,a_1,...,s_i,a_i]\)
\ENDWHILE
\end{algorithmic}
\end{algorithm}

\section{Simulation and Results}
\subsection{Simulation Configuration and  Training Tricks}
Without loss of generality, 24 users and 2 UAV-BSs are configured in the UAV network. In the simulation, the number of users served by the UAV-BSs can be changed flexibly to depict the departure and arrival of users.  In particular, users are uniformly distributed in a square area with a size of 800m × 800m at every time slot. Other parameters are list in Table I.

The actor DNN consist of an input layer with \(K \times 3 + P \times 2  = 76\) neurons, 4 fully connected hidden layers and an output layer with 4 neurons. In particular, the 4 fully connected layers have 256, 128, 64 and 16 neurons, respectively. The activation function of the output layer of the actor DNN is Tanh function, and the activation functions of the hidden layers and output layers of the actor DNN  are Relu function. The Adam optimization algorithm is used with the learning rate 0.0001. The critic DNN is composed of an input layer with \(K \times 3 + P \times 4 = 81\) neurons, the 4 fully connected layers with 256, 128, 64 and 16 neurons, respectively, and the output layer with 1 neuron. The target actor and critic DNN are updated every \(L = 200\) epoch. One episode has 800 epochs. 5000 episodes are used to train the actor and critic DNNs.

To help actor and critic DNNs converge faster, we consider the change of system throughput without  interference as the immediate reward  to pretrain the actor and critic DNNs in the early stage of the training process. The pretraining makes UAV-BSs to move to places with high user density and get many samples in which immediate rewards are diverse. Batch normalization technique is used to prevent actor and critic DNNs from overfitting.

\begin{table}[!htbp]
\caption{Simulation Configuration}
\begin{center}
\scalebox{0.8}{
\begin{tabular}{|c|c|c|c|}
\hline
\textbf{parameters} & \textbf{{values}}& \textbf{{parameters}}& \textbf{{values}} \\
\hline
Height of UAV-BS $H$ & 100m& UAV-BS transmit power $P_V$ & 1W \\
\hline
Noise power density $\sigma^2$ & -174dBm & Discounted factor $\gamma$  & 0.9 \\
\hline
$\mu_{LOS}$ & 1dB & $\eta_{NLOS}$ & 0dB \\
\hline
$\Gamma$ & 2.5bps/Hz & $\delta$   & 250m \\
\hline
Carrier frequency $f$ & 2GHz & Total bandwidth  & 20MHz \\
\hline
$B$ & 0.136 & $C$  & 11.95 \\
\hline
Critic learn rate & 0.0001  & Actor learn rate  & 0.0001 \\
\hline
Batch size & 64 & Buffer capacity  & $1\times 10^7$ \\
\hline
Epoch & 800 & Episode   & 5000 \\
\hline
\end{tabular}}
\label{tab1}
\end{center}
\end{table}

\subsection{Result and Performance Evaluating}
In this part, the simulation results are presented to evaluate the performance of the proposed AC-based UAV-BS deployment method.

 The throughput performances of the proposed AC-based UAV-BS deployment method, the sequential least-squares programming (SLSQP), heuristic and fixed UAV-BSs methods at 100 time slots are shown in Fig.2. The classical  annealing algorithm in heuristic algorithms is implemented for comparison. Compared with the throughput of heuristic algorithm, the proposed Algorithm 1 achieves a better throughput in the 78\% time slots of simulation period for UAV networks. The MINP problem is solved by the SLSQP method when the MINP problem is considered simply as a quadratic nonlinear programming problem. Compared with the throughput of the SLSQP method, the proposed Algorithm 1 achieves a better throughput in the 84\% time slots of simulation period for UAV networks. For the fixed method, the throughput is lower than the proposed method at every time slot. Besides, under the same hardware and operating system condition, the solution time of the proposed  method is 1439.93 seconds, significantly less than the 1916.90 seconds of the heuristic algorithm.
 \begin{figure}[htbp]
 \centerline{\includegraphics[width=0.3\textwidth,height=0.2\textwidth]{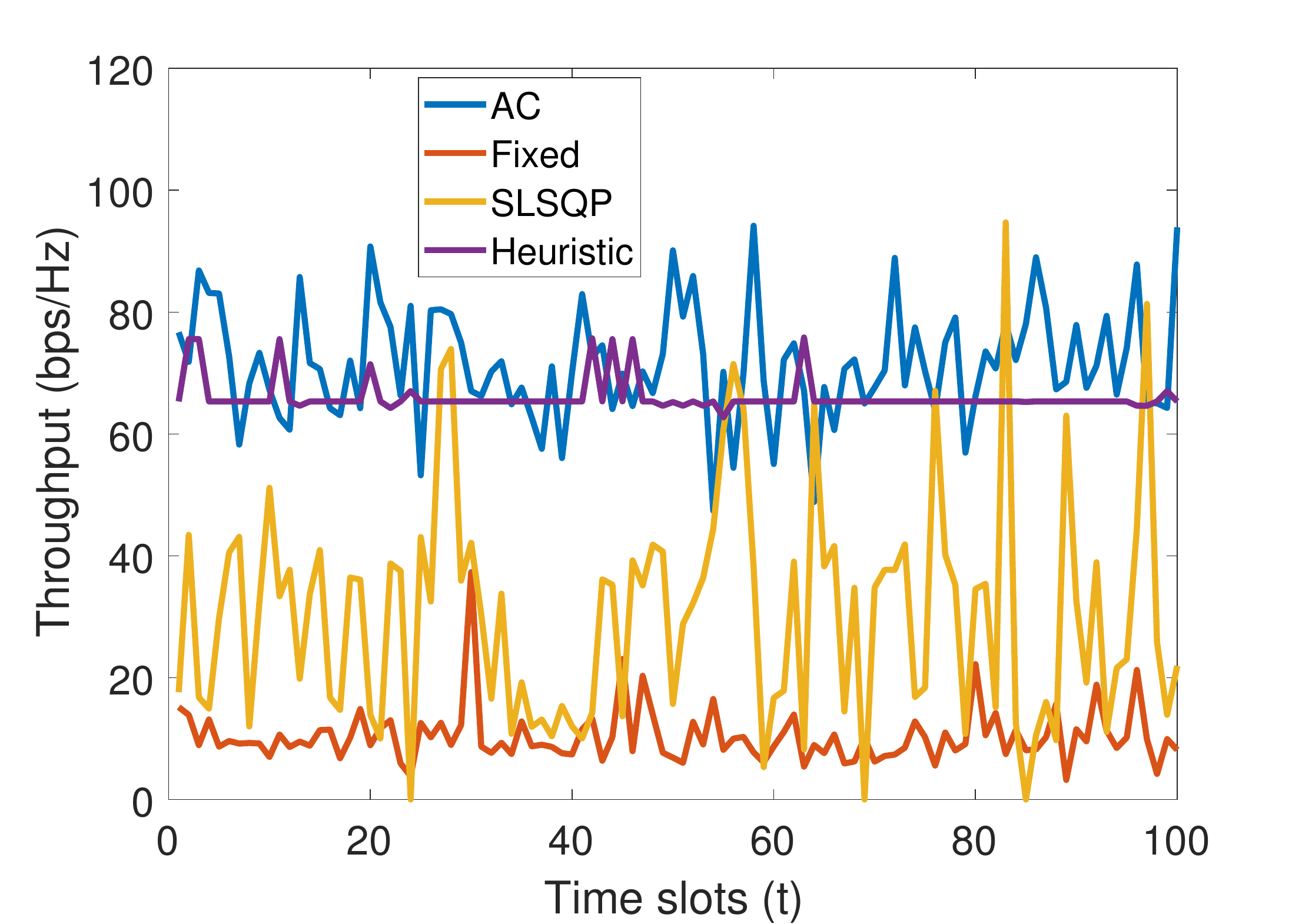}}
 \caption{Throughput of  UAV networks.}
 \label{fig}
 \end{figure}

 The long-term average throughput performances with users in the Gaussian and Uniform  distribution is shown in Fig.3. Compared with the heuristic algorithm, the
proposed method can improve the long-term average throughput by 7\% and 27\%  when the user locations are in 2-D uniform distribution and 2-D Gaussian distribution, respectively. Under the condition that the user locations are in the gaussian distribution, the  UAV-BS agents do not just tend to be in places with dense crowds, and also  have learned to cooperate with each other to avoid the interference by fine-tuning their positions. Fig.3. indicates the proposed method has good generalization. As shown in Fig.4, the throughput performance of the proposed method increases as the user density increase. Compared with the heuristic algorithm, the proposed method can maximally improve the throughput by 43.4\% when the user density varies.
 \begin{figure}[htbp]
 \centerline{\includegraphics[width=0.3\textwidth,height=0.2\textwidth]{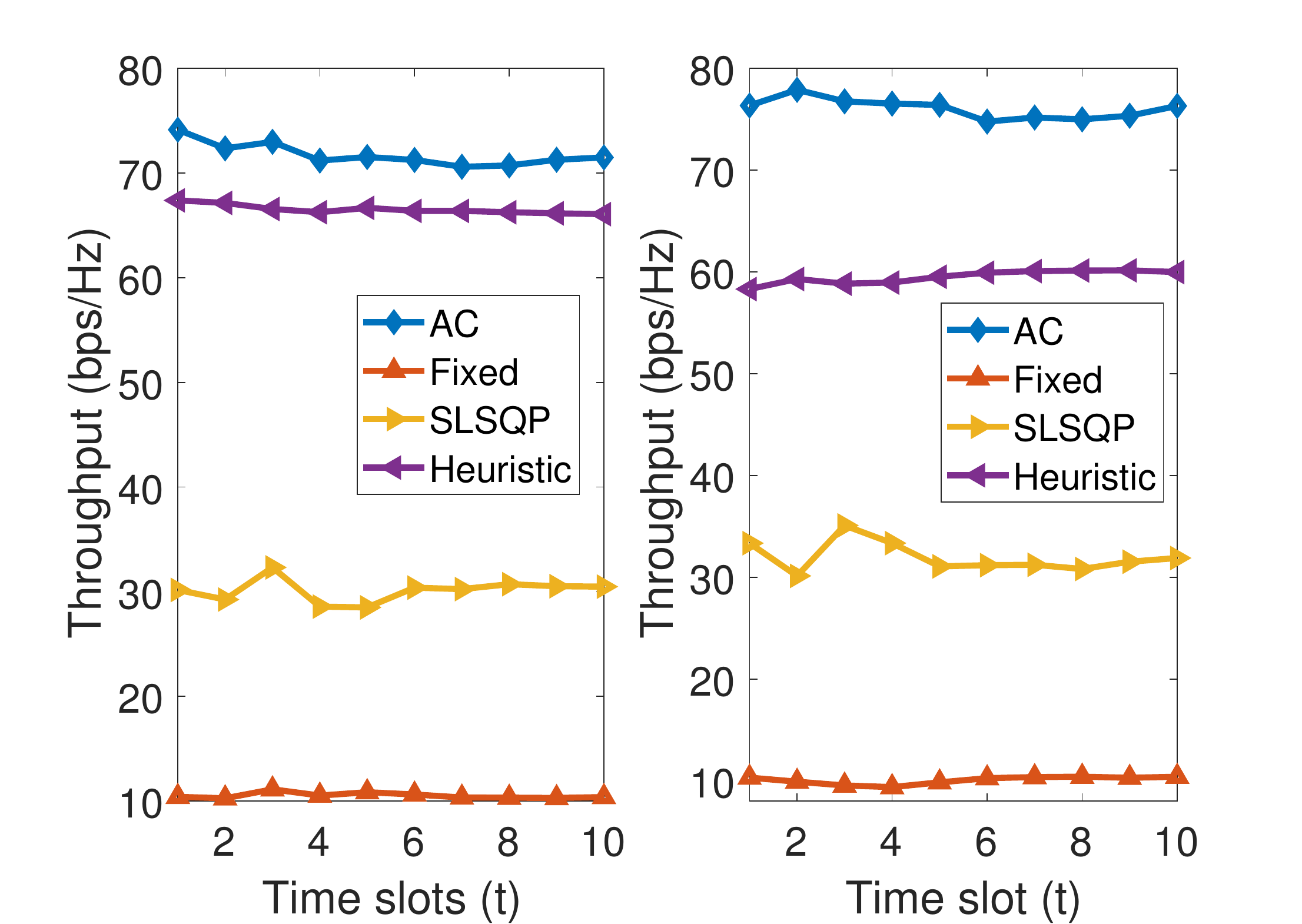}}
 \caption{Long-term average throughput of UAV networks with users in uniform (left) and Gaussian (right) distribution.}
 \label{fig}
 \end{figure}

 \begin{figure}[htbp]
 \centerline{\includegraphics[width=0.30\textwidth,height=0.2\textwidth]{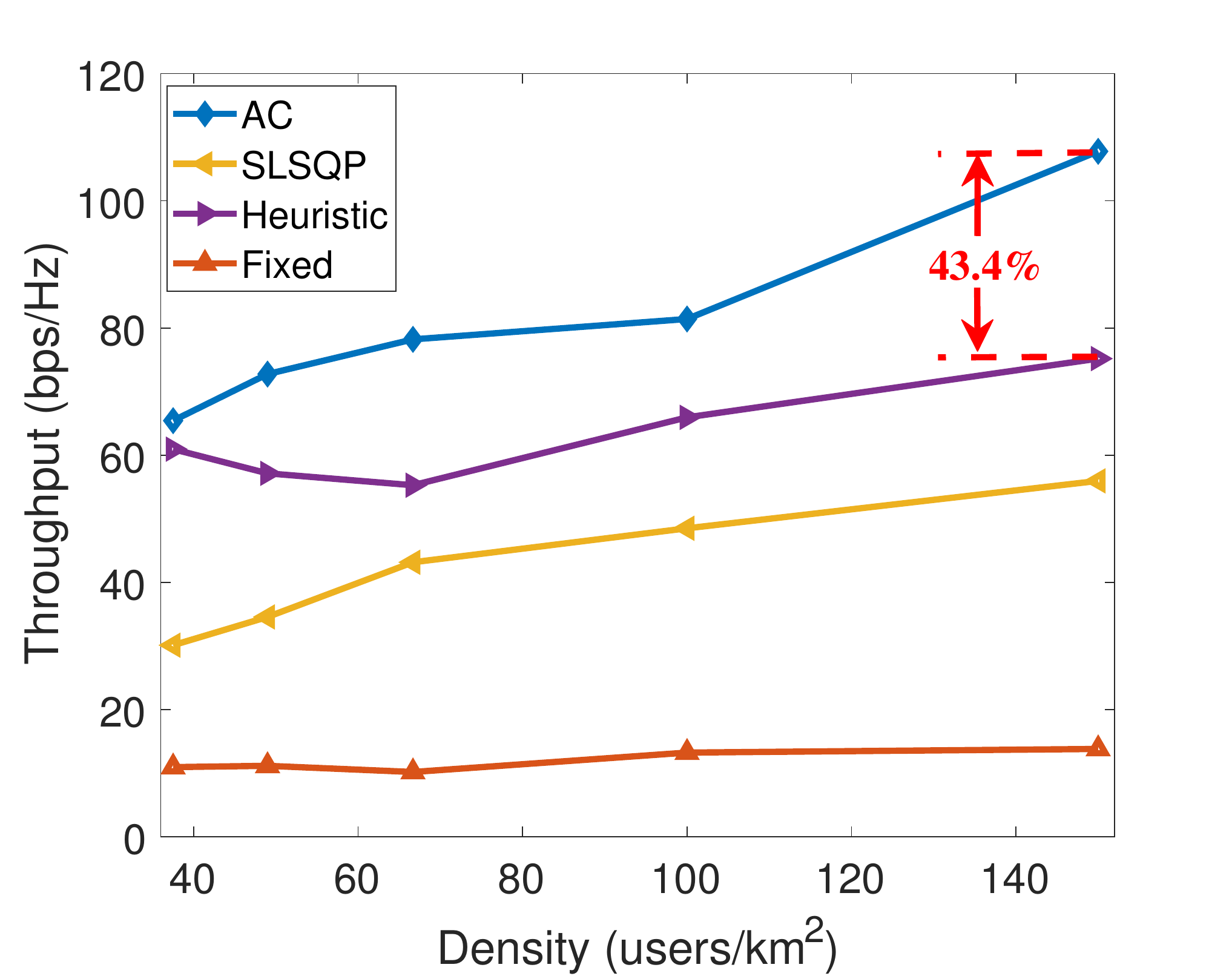}}
 \caption{Throughput of UAV networks with respect to the user density}
 \label{fig}
 \end{figure}{}

\section {Conclusion}
In this article, an AC-based DRL method for UAV-BS deployment  has been proposed to improve the throughput of  UAV networks in dynamic environments. The UAV deployment optimization is formulated as a time-varying MINP problem. To solve this problem, the process of finding the optimal UAV-BS positions is modeled as a MDP. The AC-based DRL method is used to search for the ideal UAV-BS deployment policy function. The simulation results show that the proposed  method achieves 27\% increase in the long-term average throughput for UAV networks and 24\% decrease in solution time as compared with the heuristic algorithm in UAV deployments with dynamic environments.

\section {Acknowledgment}
The authors would like to acknowledge the support from National Key R\&D Program of China (2017YFE0121600).
\bibliographystyle{IEEEtran}
\bibliography{references}

\end{document}